\documentclass[nofootinbib,twocolumn,aps,letterpaper,superscriptaddress,showpacs]{revtex4}
\usepackage{amsmath}
\usepackage{amsfonts}
\usepackage[dvips]{graphicx}
\begin{document}

\newcommand\be{\begin{equation}}
\newcommand\ee{\end{equation}}
\newcommand\bea{\begin{eqnarray}}
\newcommand\eea{\end{eqnarray}}
\newcommand\bseq{\begin{subequations}} 
\newcommand\eseq{\end{subequations}}
\newcommand\bcas{\begin{cases}}
\newcommand\ecas{\end{cases}}
\newcommand{\p}{\partial}
\newcommand{\f}{\frac}

\title{The Mixmaster Universe in a Generalized Uncertainty Principle framework}

\author{Marco Valerio Battisti}
\email{battisti@icra.it}
\affiliation{ICRA - International Center for Relativistic Astrophysics}
\affiliation{Dipartimento di Fisica (G9), Universit\`a di Roma ``Sapienza'' P.le A. Moro 5, 00185 Rome, Italy}
\author{Giovanni Montani}
\email{montani@icra.it} 
\affiliation{ICRA - International Center for Relativistic Astrophysics}
\affiliation{Dipartimento di Fisica (G9), Universit\`a di Roma ``Sapienza'' P.le A. Moro 5, 00185 Rome, Italy}
\affiliation{ENEA C.R. Frascati (Dipartimento F.P.N.), Via E. Fermi 45, 00044 Frascati, Rome, Italy}
\affiliation{ICRANET C.C. Pescara, P.le della Repubblica 10, 65100 Pescara, Italy}


\begin{abstract}
The Bianchi IX cosmological model is analyzed in a generalized uncertainty principle framework. The Arnowitt-Deser-Misner reduction of the dynamics is performed and a time-coordinate, namely the volume of the Universe, naturally arises. Such a variable is treated in the ordinary way while the anisotropies (the physical degrees of freedom) are described by a deformed Heisenberg algebra. The analysis of the model (passing through Bianchi I and II) is performed at classical level by studying the modifications induced on the symplectic geometry by the deformed algebra. We show that, the Universe can not isotropize because of the deformed Kasner dynamics, the triangular allowed domain is asymptotically stationary with respect to the particle (Universe) and its bounces against the walls are not interrupted by the deformed effects. Furthermore, no reflection law can be in general obtained since the Bianchi II model is no longer analytically integrable. This way, the deformed Mixmaster Universe can be still considered a chaotic system.
\end{abstract}


\maketitle 

\section{Introduction}

The existence of a fundamental scale, by which the continuum space-time picture that we have used from our experience at large scales probably breaks down, may be taken as a general feature of any quantum theory of gravity (for a review see \cite{gar}). An intuitive approach to introduce such a cut-off is based on deforming the canonical uncertainty relations leading to the so-called generalized uncertainty principle\footnote{Over the paper we adopt units such that $\hbar=c=16\pi G=1$.} (GUP)
\be\label{gup}
\Delta q \Delta p\geq \f 1 2\left(1+\beta (\Delta p)^2+\beta \langle{\bf p}\rangle^2\right),
\ee  
where $\beta>0$ is a deformation parameter such that for $\beta=0$ the ordinary relation is recovered. The uncertainty principle (\ref{gup}) has appeared in perturbative string theory \cite{String}, considerations on the proprieties of black holes \cite{Mag} and de Sitter space \cite{Sny}. From the string theory point of view, a minimal observable length is a consequence of the fact that strings can not probe distances below the string scale. The relation (\ref{gup}) implies a finite minimal uncertainty in the position $\Delta q_0=\sqrt\beta$ and therefore this approach entails a minimal scale in the quantum framework. However, the cut-off predicted by the GUP is, by its nature, different from the minimal length predicted by other approaches, for example the minimal eigenvalue of the geometric operators in loop quantum gravity \cite{lqg}. Recently, such an approach received notable interest and a wide work has been made on this field in a large variety of directions (see for example \cite{GUP} and the references therein). The generalized uncertainty principle (\ref{gup}) can be immediately reproduced modifying the canonical Heisenberg algebra by the following one \cite{Kem1,Kem2}
\be\label{modal}
[{\bf q},{\bf p}]=i(1+\beta{\bf {p}}^2). 
\ee 
Although such a deformed commutation relation, differently from the GUP itself, has not been so far derived directly from string theory, it represents a possible way in which certain features of a more fundamental theory may manifest themselves in some mechanical models.

In this work we analyze the Bianchi IX cosmological model (the Mixmaster Universe) in the GUP framework. This study improves a research line of ours which is centered in the investigation of cosmological models with a minimal scale \cite{BM07c,BM07a,BM07b}. The Friedmann-Robertson-Walker Universe filled with a scalar field and the Taub model have been analyzed in previous works. In the first case \cite{BM07a}, the big-bang singularity appears to be probabilistically removed but no evidences for a big-bounce (as predicted by the loop approach \cite{APS}) arise. In the second case \cite{BM07b} also, the Universe is singularity-free and furthermore the GUP wave packets provide the right behavior in the establishment of a quasi-isotropic configuration for the model. 

The Bianchi IX model, together with Bianchi VIII, is the most general homogeneous model and its physical relevance relies on the fact that it represents a {\it general} solution of the Einstein equations toward the singularity \cite{BKL}. In fact, via the Belinski-Khalatnikov-Lifshitz (BKL) scenario, when the cosmological singularity is approached in the context of a generic inhomogeneous framework, the spatial points (causal horizons) dynamically decouple and each of them evolves independently as a Bianchi IX model \cite{BKL}. The approach to the singularity of the Mixmaster model is described by a particle in two dimensions (the two physical degree of freedom of the Universe, i.e. the anisotropies) moving in a potential having exponential walls bounding a triangle \cite{Mis}. Such a particle is reflected by the walls and the dynamics appears to be chaotic \cite{chaos}. Such a model has been then used to describe the (classical) physics near the cosmological singularity.

The application of the GUP framework in quantum cosmology is well-motivated. By the minisuperspace reduction, a genuine quantum field theory (quantum general relativity) reduces to a quantum mechanical system (homogeneous quantum cosmology). As well-known, the homogeneous models, in the vacuum case, are characterized by only three degrees of freedom and therefore they are nothing but three-dimensional mechanical systems. In this respect, the GUP approach to quantum cosmology appears to be physically grounded since it can be reproduced modifying the canonical Heisenberg algebra. 

The Bianchi IX model will be studied in the context of the Arnowitt-Deser-Misner (ADM) reduction of the dynamics (for a review see \cite{rev}). Such a representation allows us to regard one variable, mainly the Universe volume, as a time for the dynamics. This model will be described by the motion of a two-dimensional particle in a triangular allowed domain. These variables, describing the physical degrees of freedom of the system, will be treated in the GUP formalism, while the time-variable in a canonical way. To perform the analysis two necessary steps, i.e. the study of the Bianchi I and II cosmological models, are necessary. The main results we obtain are in order. (i) The Bianchi I dynamics is still Kasner-like but is deeply modified since the GUP effects act in an opposite way with respect to a massless scalar field. Moreover, the deformed particle (Universe) moves faster than the ordinary case and when the Universe shrinks toward the singularity, the distances can contract along one direction while growing along the other two, i.e. two negative Kasner indices are allowed. (ii) The Bianchi II model is no longer analytically integrable and therefore no BKL map can be obtained. In other words, a relation which describes the details of the bounce of the particle against the potential walls can not be analytically found. (iii) The potential walls of Bianchi IX become stationary with respect to the particle when its momentum is of the same order of the cut-off. The triangular domain is ``dynamically closer'' than the standard one and no way for the particle to escape from the bounces arises. We conclude that the deformed evolution of the Mixmaster Universe is still chaotic.

The paper is organized as follows. In Section II the Bianchi cosmological models are reviewed and the deformed picture applied to them. Section III and IV are devoted to the analysis of the Bianchi I and II models in the GUP scheme, respectively. In Section V the deformed Bianchi IX model is investigated. Concluding remarks follow.

\section{Deformed Bianchi models}

In this Section we discuss how the equations of motion for the Bianchi models are modified by a minimal cut-off on the anisotropies. We analyze the deformations induced on the (reduced) phase space by a generalization of (\ref{modal}) in which both the two degrees of freedom of the Universe have a non-zero minimal uncertainty.    

The Bianchi Universes are spatially homogeneous cosmological models such that the symmetry group acts simply transitively\footnote{Let $G$ a Lie group, $G$ is said to act simply transitively on the spatial manifold $\Sigma$ if, for all $p,q\in\Sigma$, there is a unique element $g\in G$ such that $g(p)=q$.} on each spatial manifold \cite{RS}. The dynamics of these models is summarized in the scalar constraint which, in the Misner scheme \cite{Mis}, reads
\be\label{scacon} 
H=-p_\alpha^2+p_+^2+p_-^2+e^{4\alpha}V(\gamma_\pm)=0,
\ee
where the lapse function $N=N(t)$ has been fixed by the time gauge $\dot\alpha=1$ as $N=-e^{3\alpha}/2p_\alpha$. The variable $\alpha=\alpha(t)$ describes the isotropic expansion of the Universe while its shape changes (the anisotropies) are determinated via $\gamma_\pm=\gamma_\pm(t)$. Therefore, homogeneity reduces the phase space of general relativity to six dimensions. In this framework the cosmological singularity appears for $\alpha\rightarrow-\infty$ and the differences between the Bianchi models are summarized in the potential term $V(\gamma_\pm)$ which is related to the three-dimensional scalar of curvature. 

To describe the time evolution of the models a choice of time has to be performed. As well-known \cite{Ish} in general relativity it is possible to trace the dynamics in a relational way (with respect to an other field) or with respect to an internal time which is constructed from phase space variables. The ADM reduction of the dynamics relies on the idea to solve the scalar constraint with respect to a suitably chosen momentum. This way, we obtain an effective Hamiltonian which depends only on the physical degrees of freedom of the system. Since the volume $\mathcal V$ of the Universe is $\mathcal V\propto e^{3\alpha}$, the variable $\alpha$ can be regarded as a good clock for the evolution and therefore the ADM picture arises as soon as the constraint (\ref{scacon}) is solved with respect to $p_\alpha$. Explicitly, we obtain
\be\label{admham}
-p_\alpha=\mathcal H=\left(p_+^2+p_-^2+e^{4\alpha}V(\gamma_\pm)\right)^{1/2},
\ee 
where $\mathcal H$ is a time-dependent Hamiltonian from which is possible to extract, for a given symplectic structure, all the dynamical informations about the homogeneous cosmological models. 

Let us now analyze the modifications induced on the phase space by the GUP approach. In particular, we consider the $N$-dimensional generalization of the relation (\ref{modal}) as \cite{Kem2}
\be
[{\bf q}_i,{\bf p}_j]=i\delta_{ij}(1+\beta{\bf p}^2)+i\beta'{\bf p}_i{\bf p}_j, \qquad {\bf p}^2={\bf p}_i{\bf p}^i,
\ee
$\beta'>0$ being a new parameter. Furthermore, assuming that the translation group is not deformed, i.e. $[{\bf p}_i,{\bf p}_j]=0$, the commutation relations among the coordinates are almost uniquely determined by the Jacobi identity. The deformed classical dynamics is thus summarized in the modified symplectic geometry arising from the classical limit of the quantum-mechanical commutators, as soon as the parameters $\beta$ and $\beta'$ are regarded as independent constants with respect to $\hbar$. Therefore, the phase space algebra we consider is the one in which the fundamental Poisson brackets are \cite{Ben} 
\bea\label{defal}
\{q_i,p_j\}&=&\delta_{ij}(1+\beta p^2)+\beta'p_ip_j,\\ \nonumber
\{p_i,p_j\}&=&0,\\ \nonumber
\{q_i,q_j\}&=&\f{(2\beta-\beta')+(2\beta+\beta')\beta p^2}{1+\beta p^2}(p_iq_j-p_jq_i).
\eea  
From a string theory point of view, keeping the parameters $\beta$ and $\beta'$ fixed as $\hbar\rightarrow0$ corresponds to keeping the string momentum scale fixed while the string length scale shrinks to zero \cite{String}. In order to obtain the deformed Poisson bracket, some natural requirements have to be considered. It must posses the same proprieties as the quantum mechanical commutator, i.e. it has to be anti-symmetric, bilinear and satisfy the Leibniz rules as well as the Jacobi identity. This way, the Poisson bracket for any phase space function reads
\be
\{F,G\}=\left(\f{\p F}{\p q_i}\f{\p G}{\p p_j}-\f{\p F}{\p p_i}\f{\p G}{\p q_j}\right)\{q_i,p_j\}+\f{\p F}{\p q_i}\f{\p G}{\p q_j}\{q_i,q_j\}.
\ee
It is worth noting, that for $\beta'=2\beta$ the coordinates $q_i$ become commutative up to higher order corrections, i.e. $\{q_i,q_j\}=0+\mathcal O(\beta^2)$ and the isotropic minimal uncertainty in position reads $\Delta q_0=2\sqrt\beta$. This can be considered a preferred choice of parameters and from now on we analyze this case. However, although we neglect terms like $\mathcal O(\beta^2)$, the case in which $\beta p^2\gg1$ is allowed since in such a framework no restrictions on the $p$-domain arise, i.e. $p\in\mathbb R$.

The deformed classical dynamics of the Bianchi models can be obtained from the symplectic algebra (\ref{defal}) for $\beta'=2\beta$. The time evolution of the anisotropies and momenta, with respect to the ADM Hamiltonian (\ref{admham}), is thus given by ($i,j=\pm$)
\bea\label{defeq}
\dot \gamma_i&=&\{\gamma_i,\mathcal H\}=\f1{\mathcal H}\left[(1+\beta p^2)\delta_{ij}+2\beta p_ip_j\right]p_j,\\ \nonumber
\dot p_i&=&\{p_i,\mathcal H\}=-\f{e^{4\alpha}}{2\mathcal H}\left[(1+\beta p^2)\delta_{ij}+2\beta p_ip_j\right]\f{\p V}{\p \gamma_j},
\eea 
where the dot denotes differentiation with respect to the time variable $\alpha$ and $p^2=p_+^2+p_-^2$. These are the deformed equations of motion for the homogeneous Universes and the ordinary ones are recovered in the $\beta=0$ case. In what follow such a dynamics for the Bianchi I, II and IX models will be investigated in detail.

\section{Deformed Bianchi I model}

The Bianchi I model is the simplest homogeneous model and describes a Universe with flat space sections \cite{rev,RS}. Its line element is invariant under the group of three-dimensional translations and therefore the spatial Cauchy surfaces can be identified with $\mathbb R^3$. This Universe contains as a special case the flat FRW model which is obtained as soon as the isotropy condition is taken into account. In the above scheme, this Universe corresponds to the case $V(\gamma_\pm)=0$ and thus, from the Hamiltonian (\ref{admham}), it is described by a two-dimensional free particle (more precisely a massless scalar relativistic particle). The deformed equations of motion (\ref{defeq}) are immediately solved by
\be
\dot\gamma_\pm=C_\pm(\beta), \qquad \dot p_\pm=0,
\ee
$C(\beta)$ being a function of $\beta$. Therefore, the solution is Kasner-like. The velocity of the particle (Universe), however, is modified by the deformed geometry and, from the first equation of (\ref{defeq}), it reads
\be\label{anivel}
\dot\gamma^2=\dot\gamma_+^2+\dot\gamma_-^2=\f{p^2}{\mathcal H^2}\left(1+6\mu+9\mu^2\right)=1+6\mu+9\mu^2,
\ee 
where $\mu=\beta p^2$. In the last step we have used the fact the for the Binchi I model the ADM Hamiltonian (\ref{admham}) is given by $\mathcal H^2=p^2=$ const. As expected, for $\beta\rightarrow0$ (or better when $\mu\ll1$), the standard Kasner velocity $\dot\gamma^2=1$ is recovered. The effects of an anisotropies cut-off imply that the particle moves faster than the ordinary case. 

Let us now analyze how the Kasner behavior is modified by the deformed framework. As well-known \cite{rev,RS}, the Kasner solution is such that the spatial metric reads 
\be
dl^2=t^{2s_1}dx_1^2+t^{2s_2}dx_2^2+t^{2s_3}dx_3^2,
\ee
where $s_1, s_2, s_3$ are the so-called Kasner indices satisfying the relations $s_1+s_2+s_3=1$ and $s_1^2+s_2^2+s_3^2=1$. Only one of them is independent and except for the case $(0,0,1)$ and $(-1/3,2/3,2/3)$, such indices are never equal, but one negative and two positive. It is worth noting that the first Kasner-relation arises from the arbitrariness in choosing the tetrads, and thus is still valid in the deformed approach, while the second one is directly related to the anisotropy velocity $\dot\gamma$ by the equations \cite{rev}
\be
\dot\gamma_+=\f12(1-3s_3), \qquad \dot\gamma_-=\f{\sqrt3}2(s_1-s_2).
\ee
From formula (\ref{anivel}), the second Kasner-relation is then deformed as
\be\label{kasrel}
s_1^2+s_2^2+s_3^2=1+4\mu+6\mu^2,
\ee
and, as usual, for $\beta=0$ the standard one is recovered. Two remarks are in order. (i) From this equation, it is easy to verify that the GUP acts in an opposite way as a massless scalar field (or alternatively stiff-fluid with pressure equal to density) in the standard model. In the ordinary case, a massless scalar field allows only a finite number of oscillations in Bianchi IX before the evolution is changed into a state in which all directions shrink monotonically to zero as the curvature singularity is reached. In this case the chaotic behavior of the Mixmaster is tamed \cite{scafield}. On the other hand, in the GUP framework, all the terms on the right hand side of (\ref{kasrel}) are positive and it means that the Universe cannot isotropize, i.e. it can not reach the stage such that the Kasner indices are equal. (ii) For every non-zero $\mu$, the modifications induced on the standard Kasner behavior are significant since two indices can be negative at the same time. In other words, as the volume of the Universe contracts toward the classical singularity, distances can shrink along one direction and grow along the other two. In the ordinary case the contraction is along two directions. Therefore, even if a ``quasi-standard'' regime ($\mu\ll1$) is addressed, the Kasner dynamics is deeply modified by such an approach. 

\section{Deformed Bianchi II Model}

Let us now investigate the dynamics of Bianchi II in the framework of the deformed phase space discussed above. This model is a fundamental step toward the Bianchi IX one. It represents a bridge from the flat homogeneous model (Bianchi I) and the Mixmaster Universe (Bianchi IX). Its dynamics is the one of a two-dimensional particle bouncing against a single wall. More precisely, it corresponds to the Mixmaster dynamics when only one of the three equivalent potential walls is taken into account \cite{rev,RS}. The main features of Bianchi IX, as the BKL map, are obtained considering such a simplified model since it is, in the ordinary framework, an integrable system differently from Bianchi IX itself \cite{Misq}.  

In the Hamiltonian framework, Bianchi II is the homogeneous model for which the potential term is given by $V(\gamma_\pm)=e^{-8\gamma_+}$ and such an expression can be obtained from the one of Bianchi IX in a given asymptotic region. The ADM Hamiltonian (\ref{admham}) in this case reads
\be\label{ham2}
\mathcal H=\left(p_+^2+p_-^2+e^{4(\alpha-2\gamma_+)}\right)^{1/2}.
\ee
Our aim is to describe the bounce of the particle (Universe) against the potential wall in the GUP scheme. A fundamental difference with respect to the ordinary case is that $\mathcal H$ is no longer a constant of motion near the classical singularity ($\alpha\rightarrow-\infty$). In the undeformed scheme the anisotropy velocities are simply given by $\dot\gamma_\pm=p_\pm/\mathcal H$ and, from (\ref{ham2}), one immediately obtains $\dot{\mathcal H}=0$ for $\alpha\rightarrow-\infty$. 

In our scheme such a feature is modified by the deformation terms and, in particular, by the velocity relation (\ref{anivel}) which replaces the ordinary one $\dot\gamma^2=p^2/\mathcal H^2$. The equation $\dot{\mathcal H}=\p_\alpha\mathcal H$ in the deformed framework gives
\be\label{devh}
\f\p{\p\alpha}(\ln\mathcal H^2)=4\left(1-\f{p^2(\dot\gamma)}{\mathcal H^2}\right)\neq4\left(1-\dot\gamma^2\right),
\ee
and is no longer equal to zero. Of course, with $p(\dot\gamma)$ we have indicated the solution of the velocity equation (\ref{anivel}). From the above relation (\ref{devh}) it is possible to compute the velocity of the potential wall. The condition that the potential term $V(\gamma_\pm)=e^{-8\gamma_+}$ be important near the cosmological singularity is easily seen to be $e^{4(\alpha-2\gamma_+)}\simeq\mathcal H^2$. The potential (wall) velocity $\dot\gamma_w$ then reads
\be\label{velw}
\dot\gamma_+\simeq\dot\gamma_w=\f12-\f18\f\p{\p\alpha}(\ln\mathcal H^2)=\f{p^2(\dot\gamma)}{2\mathcal H^2}.
\ee
As in the ordinary case, $\gamma_w$ defines the equipotentials in the anisotropy plane, where the potential term is relevant. In the standard picture, since $\dot\gamma^2=1$, the wall velocity is equal to one half of the particle one, i.e. $\dot\gamma_w=1/2$. 

The undeformed dynamics toward the classical singularity ($\alpha\rightarrow-\infty$) is as follows. The anisotropy particle $\gamma(\alpha)$ moves with velocity $\dot\gamma=1$ except when it approaches the equipotentials $e^{4(\alpha-2\gamma_+)}\simeq\mathcal H^2$. This wall moves outward with a velocity $\dot\gamma_w=1/2$ and therefore, in a finite time interval, the particle will bounce against it. The dynamics of the particle before and after this collision is the one described by the Bianchi I model.

In the deformed case, both the particle and the potential wall move faster than the ordinary one. The main point is to establish if there exists a range in which the wall moves faster than the anisotropy particle. In fact, in such a case, the point-Universe no longer bounces against the wall and the Kasner behavior remains unaltered. From the above relations (\ref{anivel}) and (\ref{velw}), it is possible to derive the explicit form of the wall velocity, which reads (near the singularity)
\be\label{walvel}
\dot\gamma_w=\f1{36\mu}\left(-4+2^{1/3}2g^{-1/3}+2^{2/3}g^{1/3}\right),
\ee
where $g=g(\mu)$ is defined as $g=2+81\mu\dot\gamma^2+9\sqrt{\mu\dot\gamma^2(4+81\mu\dot\gamma^2)}$. We stress that, near the cosmological singularity ($\alpha\rightarrow-\infty$) we have $\mathcal H^2\simeq p^2$ and therefore the particle velocity $\dot\gamma$ is the same as in the Bianchi I case. Moreover, it is not difficult to see that for $\beta\rightarrow0$ the ordinary velocity of the potential wall is recovered and that the bounce always occurs also in the deformed scheme. 

Let us now discuss the details of the bounce. In the standard case the particle (Universe) moves twice as fast as the receding potential wall, independently of its momentum (namely its energy). In the deformed framework the particle velocity, as well as the potential velocity, depends on the anisotropy momenta and on the deformation parameter $\beta$. In this case also the particle moves faster than the wall, since the relation $\dot\gamma_w<\dot\gamma$ is always verified. Thus, a bounce takes place also in the deformed picture. Furthermore, the wall appears asymptotically stationary when the particle has a growing energy, i.e. when the region $\mu\gg1$ is investigated (we recall that $p\in\mathbb R$). In this limit, the relation $\dot\gamma_w/\dot\gamma\sim1/(6\mu)$ holds (see Fig. 1) and then, for $\mu\gg1$, no limit angle for the collision appears. More precisely, let us indicate with $\theta_i$ and $\theta_f$ the angles of incidence and of reflection for the bounce, respectively. The velocity $\dot\gamma$ is parametrized as follows \cite{Misq}: in the initial state we have $(\dot\gamma_+)_i=-\dot\gamma\cos\theta_i$, $(\dot\gamma_-)_i=\dot\gamma\sin\theta_i$ and in the final one $(\dot\gamma_+)_f=\dot\gamma\cos\theta_f$, $(\dot\gamma_-)_f=\dot\gamma\sin\theta_f$. Thus, the maximum angle in order the bounce against the wall to occur is given by $|\theta_i|<|\theta_{\max}|=\cos^{-1}(\dot\gamma_w/\dot\gamma)$ and hence, in the asymptotic limit $\mu\gg1$, we have $|\theta_{\max}|=\pi/2$. In the ordinary case ($\dot\gamma_w/\dot\gamma=1/2$), the maximum incidence angle is given by $|\theta_{\max}|=\pi/3$ \cite{Misq}.
\begin{figure}
\begin{center}
\includegraphics[height=1.8in]{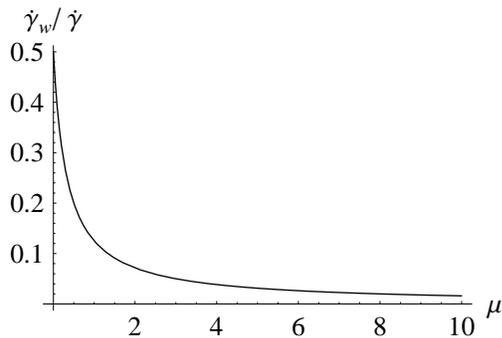}
\caption{The potential wall velocity $\dot\gamma_w$ with respect to the particle one $\dot\gamma$ in function of $\mu=\beta p^2$. In the $\mu\rightarrow0$ limit, the ordinary behavior $\dot\gamma_w/\dot\gamma=1/2$ is recovered.} 
\end{center}
\end{figure} 

The next step would be to obtain the reflection law (the BKL map) which connects the initial ($\theta_i$) and final ($\theta_f$) between the particle-velocity and the wall. In order to integrate the model (\ref{ham2}), we have to recover two first integrals of motion. In the ordinary case, one of them is immediately found in $p_-$. On the other hand, in the deformed framework, the equations of motion (\ref{defeq}) become coupled by $\beta$-terms and in this case
\be
\dot p_-=\f{8\beta}{\mathcal H} p_+p_-e^{4(\alpha-2\gamma_+)},
\ee
which no longer vanishes unless the $\beta=0$ case is considered. In the undeformed scheme, the other constant of motion can be recovered by a linear combination of $p_+$ and $\mathcal H$, in particular by $\Omega=\mathcal H-p_+/2$. From such constants of motion, is not difficult to obtain the reflection law as $2(\sin\theta_f-\sin\theta_i)=\sin(\theta_i+\theta_f)$ \cite{Misq}. Contrarily, in the deformed picture, the remaining equations of motion read
\bea
\dot p_+&=&\f4{\mathcal H}e^{4(\alpha-2\gamma_+)}(1+3\beta p_+^2+\beta p_-^2),\\ \nonumber
\dot{\mathcal H}&=&\f2{\mathcal H}e^{4(\alpha-2\gamma_+)}, 
\eea 
and $\Omega$ is a constant of motion if and only if the $\beta=0$ case is taken into account. This way, differently from the standard case, the Bianchi II model in such a framework appears to be a non (analytically) integrable system. As a matter of fact no first integrals of motion can be recovered, i.e. an equation which gives $\theta_f$ in terms of $\theta_i$ can not be in general obtained. 

To gain insight onto the physical features of the model, we can consider the special cases for which a reflection law can be fixed. We analyze the different situations in which $p_+\gg p_-$ ($p_-\gg p_+$) and $p_+=p_-$. In the first case\footnote{The complementary case ($p_-\gg p_+$) is qualitatively the same.} ($p_+\gg p_-$ corresponds to $|(\dot\gamma_+)_i|\gg|(\dot\gamma_-)_i|$), two constants of motion can be obtained and read
\bea\label{conmot}
\Omega&=&\mathcal H-\f1{2\sqrt{3\beta}}\tan^{-1}(\sqrt{3\beta}p_+),\\ \nonumber
K&=&\f13\ln(1+3\beta p_+^2)-\ln(p_-).
\eea   
and as $\mu\rightarrow0$ the ordinary framework is recovered. By the use of the equations of motion $\dot\gamma_\pm$ given by (\ref{defeq}), it is possible to obtain the required reflection law between $\theta_i$ and $\theta_f$. An interesting feature appears as soon as the ultra-deformed case is considered, i.e. when $\mu\gg1$. In such a range, the two constants of motion (\ref{conmot}) become $\Omega\simeq\mathcal H$ and $K\simeq\ln{(3\beta p_+^2)}/3$ and thus the reflection law is given by $|\theta_i|=|\theta_f|$ as in the usual framework. 

In the second peculiar case ($p_+=p_-$), the two first integrals of motion are
\bea
\Omega&=&\mathcal H-\f1{4\sqrt\beta}\tan^{-1}(2\sqrt\beta p_+),\\ \nonumber
K&=&\f{p_+}2+\f1{4\sqrt\beta}\tan^{-1}(2\sqrt\beta p_+).
\eea
Also in this case, considering equations (\ref{defeq}) the map $\theta_f=\theta_f(\theta_i)$ can be obtained and in the ultra-deformed regime ($\mu\gg1$) it reads $|\theta_i|=|\theta_f|$ as in the ordinary scheme.

Let us now summarize the effects of the deformed framework on the dynamics of the Bianchi II model. The main difference with respect to the ordinary picture is that such a model is no longer integrable and therefore no reflection map can be in general inferred. It can be obtained only in few peculiar cases. The other important feature of our model is that the potential wall becomes stationary, with respect to the $\gamma$-particle, in the asymptotic regime $\mu\gg1$ (see Fig. 1). Therefore, when the momenta of the particle (Universe) reaches the cut-off value $\beta$, its bounce against the wall is improved in the sense that no longer maximum limit angle appears. 

\section{Deformed Mixmaster Universe}

In this section we describe the deformed Mixmaster Universe. We analyze the deformed phase space of the Bianchi IX cosmological model in agreement with the previous discussion. The Bianchi IX geometry is invariant under the three-dimensional rotation group and therefore the space-time manifold can be topologically written as $\mathcal M=\mathbb R\otimes SO(3)$ \cite{rev,RS}. Thus, this Universe is the generalization of the closed FRW model when the isotropy hypothesis is relaxed. In the Hamiltonian formulation, it appears as soon as the potential term $V(\gamma_\pm)=e^{4(\gamma_++\sqrt3\gamma_-)}+e^{4(\gamma_+-\sqrt3\gamma_-)}+e^{-8\gamma_+}$ in (\ref{admham}) is taken into account. Such a potential delimits a triangular domain in the $\gamma$-plane where the dynamics is restricted \cite{Mis,Misq}. As well known, the evolution of the Mixmaster Universe is that of a two-dimensional particle bouncing infinite times against three walls which rise steeply toward the singularity. In particular, between two succeeding bounces the system is described by the Kasner evolution and the permutations of the expanding-contracting directions is given by the BKL map \cite{BKL}. Such a dynamics is also chaotic \cite{chaos}. 

From the analysis of the deformed Bianchi I and II models we know several features of the deformed Mixmaster Universe. Inside the closed domain the $\gamma$-particle moves freely and therefore its velocity is given by the formula (\ref{anivel}). The Bianchi II model, appearing as soon as only one of the three equivalent walls is taken into account, is recovered when the asymptotic region $\gamma_+\rightarrow-\infty$, $|\gamma_-|<-\sqrt3\gamma_+$ of the Bianchi IX model is considered. The velocity of the potential walls is then the same as previously computed, see equation (\ref{walvel}). In Bianchi II, because of the presence of a single potential wall, the particle performs only one bounce and then it runs freely toward the singularity. Differently, in the Bianchi IX case the particle will collide infinite times against the three walls. Two conclusions on the deformed Mixmaster Universe can be inferred. 
\begin{itemize}
	\item When the ultra-deformed regime is reached ($\mu\gg1$), i.e. when the $\gamma$-particle (Universe) has the momentum bigger than the cut-off one, the triangular closed domain appears to be stationary with respect to the particle itself. This way, the bounces of such a particle are increased by the presence of deformation terms, i.e. by the non-zero minimal uncertainty in the anisotropies.
	\item No BKL map (reflection law) can be in general obtained. It arises analyzing the single bounce against a given wall of the equilateral-triangular domain and the Bianchi II model is no longer an integrable system in the deformed picture. In other words, the chaotic behavior of the Bianchi IX model is not tamed by GUP effects, i.e. the deformed Mixmaster Universe is still a chaotic system.  
\end{itemize}
We have to stress a point. The chaoticity of Bianchi IX arises from the analysis of the stochastic proprieties of such a model and in particular from the BKL map. As we have seen, no reflection law for the Bianchi II model can be obtained in the deformed framework. Therefore, since the BKL map is constructed from the reflection law of Bianchi II, no deformed map arises at all and no quantitative predictions can be made for the model. We can however use qualitative arguments to realize that the chaoticity of Bianchi IX is not tamed by the GUP effects. In the ordinary framework, Bianchi II is an integrable model which is a part of a chaotic system (Bianchi IX). On the other hand, in the deformed framework Bianchi II, which is still a part of Bianchi IX, is no longer an integrable model. The deformations induced by a minimal uncertainty make the model much more complicated and surely they are not able to cast a chaotic system in a non-chaotic one. This is the meaning when we claim that the deformed Mixmaster Universe is still a chaotic system. It is worth noting that effects due to two negative Kasner indices arise in the modified dynamics. This new issue would require additional investigation, but it seems no way related to a possible removal of the chaoticity of this model. 

It is interesting to point out the differences between our model and the loop Mixmaster dynamics \cite{Mixl}. Loop quantum cosmology \cite{lqc} is based on the discrete structure of space predicted by loop quantum gravity. When such a framework is applied to the Bianchi IX model the classical reflections of the $\gamma$-particle stop after a finite amount of time and, when the Planck scale is reached, the height of the potential walls rapidly decreases until they completely disappear. This way, the Mixmaster chaos is suppressed by (loop) quantum effects \cite{Mixl}. In the loop framework, although the analysis is performed through the ADM reduction of the dynamics as we did, all the three scale factors are quantized using the loop techniques. On the other hand, in our approach the time variable (related to the volume of the Universe) is treated in the standard way and only the two physical degrees of freedom of the Universe (the anisotropies) are considered as deformed. This makes clear the differences between these two cut-off approaches. In fact, we expect that if we implement the deformed framework to the whole phase space, modifications on the Universe volume, i.e. on the height of the potential walls, can appear. However, on the basis of \cite{bat}, to reproduce the loop phenomenology a Snyder-deformed Heisenberg algebra should be taken into account.   

\section{Concluding remarks}

In this paper we have shown the effects of a modified Heisenberg algebra, which reproduces a GUP as arises from studies on string theory \cite{String}, on the Bianchi I, II and IX cosmological models (for other analysis of low-energy-string-effective cosmological models see \cite{lowstring}). The dynamics of these Universes is analyzed in the ADM formalism by which the variable $\alpha$ (namely the volume of the Universe) is regarded as the time-coordinate for the dynamics. Such a time variable is described in the standard way. On the other hand, the two physical degrees of freedom of the Universe (the shape changes $\gamma_\pm$) are treated according to the GUP prescription, i.e. by using a deformed Heisenberg algebra. A fundamental scale is then introduced in these models by the appearance of a non-zero minimal uncertainty in the anisotropies. The analysis of the dynamics is performed at classical level taking into account the modifications induced on the phase space by the deformed algebra. In particular, the deformed dynamics of the particle as well as of the potential walls is investigated in detail. Three main conclusions can be inferred.
\begin{itemize}
	\item The velocity of the $\gamma$-particle (Universe) inside the allowed domain of the Mixmaster model grows with respect to the undeformed case. The deformation effects, acting as opposite to a stiff-matter, imply that the Universe cannot isotropize. Furthermore, although the dynamics is still Kasner-like, two negative Kasner indices are now allowed. During each Kasner era, the volume of the Universe can contract in one direction while expands in the other two.
	\item The velocity $\dot\gamma_w$ of the potential walls, bounding the triangular domain of Bianchi IX, is increased by the deformation terms. However, it no rises so much to avoid the bounces of the $\gamma$-particle against the walls, i.e. the particle bounces are not stopped by the GUP effects. As matter of fact, when the ultra-deformed regime is reached (when $\mu\gg1$) the dynamics is that of a particle which bounces against stationary walls (no maximum incidence angle appears).
	\item No BKL map (reflection law $\theta_f=\theta_f(\theta_i)$) can be in general analytically computed. In fact, such a map arises from the analysis of the Bianchi II model which is no longer analytically integrable in the deformed scheme. A non-vanishing minimal uncertainty in the anisotropies complicates so much the Mixmaster dynamics in such a way that each of its wall-side is no longer an integrable system. We can then conclude that the chaoticity of the Bianchi IX model is not tamed by the GUP effects on the Universe anisotropies. 
\end{itemize}

\end{document}